\documentclass{IOS-Book-Article}
\usepackage{mathptmx}
\usepackage{graphicx}
\graphicspath{ {images/} }

\begin{document}

\pagestyle{headings}
\def\thepage{}

\begin{frontmatter}              

\title{Towards ontology driven learning of visual concept detectors}


\author{\fnms{Sanchit} \snm{Arora}}, %
\author{\fnms{Chuck} \snm{Cho}}, %
\author{\fnms{Paul} \snm{Fitzpatrick}}, %
\author{\fnms{Fran\c{c}ois} \snm{Scharffe}
\thanks{Corresponding Author: francois.scharffe@dextro.co}}

\runningauthor{ et al.}
\address{Dextro Robotics, Inc. 101 Avenue of the Americas, New York, USA}

\begin{abstract}
The maturity of deep learning techniques has led in recent years to a breakthrough in object recognition in visual media. While for some specific benchmarks, neural techniques seem to match if not outperform human judgement, challenges are still open for detecting arbitrary concepts in arbitrary videos.
In this paper, we propose a system that combines neural techniques, a large scale visual concepts ontology, and an active learning loop, to provide on the fly model learning of arbitrary concepts. We give an overview of the system as a whole, and focus on the central role of the ontology for guiding and bootstrapping the learning of new concepts, improving the recall of concept detection, and, on the user end, providing semantic search on a library of annotated videos.
\end{abstract}

\begin{keyword}
Computer Vision\sep Ontology\sep Deep Learning \sep Neural Networks \sep Active Learning
\end{keyword}
\end{frontmatter}
\pagestyle{empty}

\section{Introduction}

In recent years, advances in neural network design and optimization, combined with dedicated computing architectures through the use of GPUs, led to a dramatic increase in the performance of computer vision systems. 

The ILSVRC-2012 challenge winners Krizhevsky et al \cite{krizhevsky} demonstrated how given a large dataset of labeled images, one could design a  neural network and train it to achieve state-of-the-art classification results. Availability of increasingly powerful hardware has enabled training of deeper and deeper neural networks \cite{VGG} with corresponding improvements in performance. Research efforts have resulted in multiple advances including novel designs \cite{reception} and their combination with existing computer vision techniques like region proposals \cite{rcnn}, optical flow \cite{opticalflow} or long short term memories \cite{lstm}. All this has accelerated the pace at which computer vision systems approach human performance.

While large systems try to learn models for an exhaustive list of visual concepts, we take the approach of incrementally learning new concepts as they appear in new videos submitted for analysis. One of the reason for taking that path is that we need to ensure that enough training material is available to learn new models. While data is in some cases available \cite{fcvid, yto, ucf101}, it is in the general case difficult to find good video training material. Also, the computing costs related to training a model for a large number of concepts at once would not be sustainable. Finally, the quality of the learned model can be improved if the learning is guided: starting learning higher level visual concepts, ie $Car$, before refining the model to learn more specific ones like $Sports\ car$, or $Convertible\ car$.

In the following section we present a system designed by following these requirements. We then focus on the role of the ontology, motivating it as a critical component of the system. At the time of writing, the system is implemented and includes a visual concepts ontology under construction, inspired by existing visual ontologies \cite{comm, vco, hollink}. This paper is thus more positioned towards motivating the described approach rather than giving experimental results on its benefits.

\section{A Virtuous Circle for Visual Concept Learning}

\label{sect:virtuous}

This section describes the Dextro system which can be considered as two parts which come together to form an active learning loop. The first part is the Dextro API which allows anyone to analyze and understand their video by providing a link. The computer vision models combined with a human in the loop system analyze the video and promptly return results. The second part is the Dextro training system which continues to collect training videos and uses human annotators to collect data for those videos which are eventually used as training data for the computer vision models to expand their visual concept classification. The system is illustrated in Figure~1.

\begin{figure}
\includegraphics[width=0.7\textwidth]{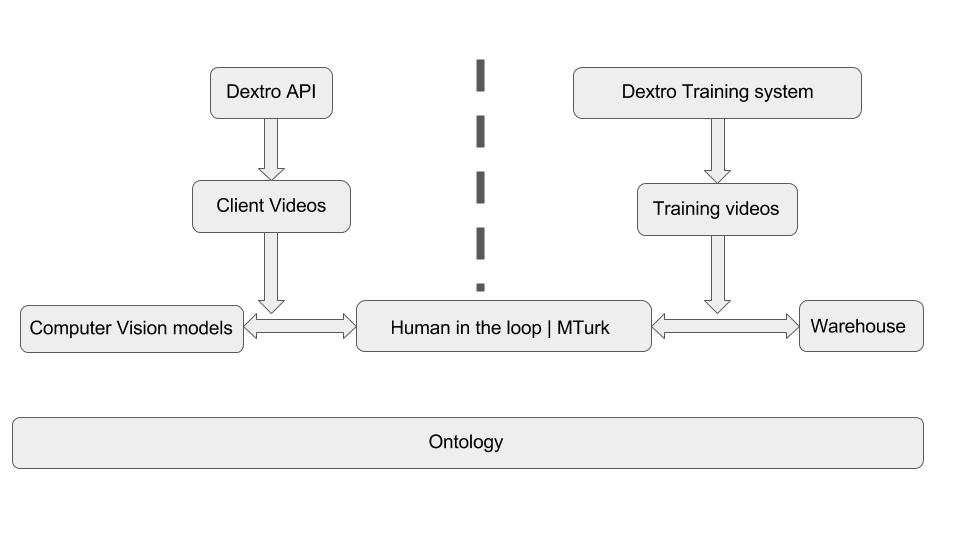}
\label{fig:dextro}
\caption{An active learning loop for video analysis}
\end{figure}

The initial computer vision system starts with a limited set of visual concepts that it is trained on. As videos are processed via the Dextro API the model is unable to confidently classify the video contents, given its limited training. Thus the video is sent to a human in the loop system where human annotators look at the video and classify it.

Instead of giving human annotators free reign on how to classify the video, we limit them to use terms from our visual ontology. This process gives us information on what type of concepts are present in the videos being sent to our API and we share that information with the Dextro training system. The training system then surfs the web for publicly available video data likely to contain the visual concepts that are in our ontology and have been most frequently used by human annotators to classify the videos.

The video data collected by the training system is verified and annotated by humans via a service like Amazon Mechanical Turk. This data is then used to retrain our computer vision system, introducing new concepts to it. Future videos with those concepts thus do not require a human for classification, completing the active learning loop.

We compare this to how humans learn about new concepts. First there is the phase of noticing something new that we do not know about. Then we involve an expert or someone more knowledgeable about the concept to help us identify and understand the same. Once we have learned the new concept we can use it in future situations.

\section{Ontology Driven Neural Model Learning}\label{sec:ontology}

As Section~\ref{sect:virtuous} illustrated, there are multiple areas where the use of a comprehensive ontology can improve the process of incrementally learning a visual concept detection model.

\textbf{Guiding the concept selection process for human annotators: } 
Human workers select visual concepts to annotate videos. Better than presenting a list of concepts, the use of an ontology will allow us to present the concept hierarchy, improving the time taken to reach the correct concept. The use of the ontology will also allow us to finely tune the level of granularity at which we want the annotations to be. If we are training on a video set in a specific domain, like car racing, then we will allow a deeper granularity in the types of e.g. $Car$, while we may want to stop at $Car$ for generalist video sets. 

\textbf{Bootstrapping model learning through the concept hierarchy: } 
When a model is learned for a concept, the concept hierarchy can be used to propagate this model to the concept’s parents, children and siblings. For example, the model for $Laptop$ can be used to bootstrap the model for $MacBook$. The bootstrapping will enable faster learning of the model and thus a lower usage of computing resources.

\textbf{Automated timeline construction: }
We can use the ontology to bootstrap from coarse human annotations of what concepts appear in a video to detailed timelines of when and where those concepts occur.  For example, if we are told a video contains a $Bison$ and no other animals, we can automatically draw on trained models for other animals such as a $Cow$ to locate the $Bison$ (just as a human might).  In general, the ontology can tell us when the models we have are sufficient to make the visual distinctions needed in a particular video by machine, and when we need more help from humans.

\textbf{Using relationships and categories to improve concepts detection: }
When a concept is detected in a video, relationships can be followed in the ontology to others concepts, indicating the potential presence of those, either in the same frame, or previously, or later in the video. For example, if a $Wheel$ is detected, and the ontology specifies that $Car\ hasPart\ Wheel$, this relation can be followed to trigger the activation of $Car$. Categorical information can also be used. For example, if a $Frying\ Pan$ is detected, other visual concepts and activities in the category $Cooking$ will be expected. An attention model can then be used to focus on features in this category.

\textbf{Semantic search on the annotated videos: } A classic application of ontology based system is to provide semantic search. For example, querying for $Computer$ will return video sequences containing $Laptop$ and $Desktop\ Computer$, even if there is no explicit model learned for the concept $Computer$. Similarly, browsing a hierarchy of concepts is a better experience than a flat list of tags.

\textbf{Bootstrapping model learning using linked data: } Thanks to the rich interconnection of public domain datasets provided in projects like DBpedia and Wikidata, we are able to link concepts to a large repository of associated media: Wikimedia Commons. This enables bootstrapping  and enriching the training set for a large number of concepts. Mappings between concepts in the ontology and linked-data resources is made possible by the ecosystem of standards and tools associated with modern ontology engineering.

\textbf{More accurate and efficient contextual model: } Contextual models exploit inter-relationships between different visual concepts, and can significantly improve classification results. For example, simple co-occurrence statistics can be constructed with training data and used at test time. In a real-world scenario where the number of concepts can be large, the co-occurrence matrix is not only expensive to construct, but also is very sparse, hence, not representative of real-world data. Ontology-aware hierarchical design of a co-occurrence matrix can remedy this.

\section{Conclusion}

While the combination of numerical and symbolic techniques may be a longer term research agenda, ontologies can bring significant improvements to visual concept detection. We have presented a system featuring an active learning loop for visual concept detection, in which the ontology plays a central role for guiding the learning process.      

The systems presented in this paper is a work in progress. The active learning neural network visual concept learning loop is in place. We are currently focusing on building and integrating a comprehensive ontology structuring the hundreds of concepts our system is currently able to detect. Our next steps will be using the ontology to perform the improvements discussed in this paper: bootstrapping model learning, and guiding the categorization process using the concept hierarchy.

\end{document}